\documentclass[journal=jpccck,manuscript=article,layout=twocolumn,psfig,pstricks]{achemso}

\setkeys{acs}{super=true}
\setkeys{acs}{articletitle=true}
\setkeys{acs}{chaptertitle=true}
\setkeys{acs}{etalmode=truncate,maxauthors=0}
\setcitestyle{super,open={},close={}}

\usepackage{amsfonts}
\usepackage{amsmath}                    
\usepackage{amssymb}                    
\usepackage[english]{babel}             
\usepackage{bm}                         
\usepackage{booktabs}                   
\usepackage{cancel}                     
\usepackage[format=plain,singlelinecheck=false,font={small,bf},labelfont=bf,labelsep=space]{caption}
\usepackage{colortbl}                   
\usepackage{csquotes}                   
\usepackage{dcolumn}                    
\usepackage{fancyhdr}                   
\usepackage{float}                      
\usepackage{geometry}                   
\usepackage{graphicx}                   
\usepackage{helvet}                     
\usepackage{hyperref}                   
\usepackage{cleveref}
\hypersetup{breaklinks=true,colorlinks=true,citecolor=blue,linkcolor=blue,filecolor=blue,urlcolor=blue}
\usepackage{listings}                   
\usepackage{lscape}                     
\usepackage{mathpazo}                   
\usepackage{mathtools}                  
\usepackage[version=4]{mhchem}          
\usepackage{microtype}                  
\usepackage{multirow}                   
\usepackage{natbib}                     
\usepackage{natmove}                    
\usepackage{pgfplots}                   
\usepackage{pstricks}                                                           
                                                        
\usepackage[fontsize=10.0pt]{scrextend}
\usepackage{setspace} 
\usepackage{siunitx}                    
\usepackage{subfigure}                  
\usepackage{tablefootnote}              
\usepackage{txfonts}                    
\usepackage[normalem]{ulem}             
\usepackage{xcolor}                     
\usepackage{xkeyval}                    
\usepackage{xspace}
\usepackage{placeins}
\mhchemoptions{textfontcommand=\sffamily} 
\mhchemoptions{mathfontcommand=\mathsf}

\pgfplotsset{compat=1.16}

\SectionsOn
\SectionNumbersOn
\AbstractOn
\newcommand{\orcid}[1]{\href{https://orcid.org/#1}{\includegraphics[width=9pt]{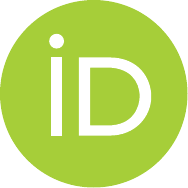}}}

\DeclareSIUnit\calorie{cal}

\author{Rafael L. H. Freire\orcid{0000-0002-4738-3120}}
\email{rafael.freire@lnnano.cnpem.br}
\affiliation[CNPEM]{Brazilian Nanotechnology National Laboratory, Campinas, SP 13083-97, Brazil}

\author{Felipe Crasto de Lima\orcid{0000-0002-2937-2620}}
\affiliation[CNPEM]{Brazilian Nanotechnology National Laboratory, Zip Code 13083-970, Campinas, SP, Brazil}
\alsoaffiliation{Ilum School of Science, CNPEM, Zip Code 13083-970, Campinas, SP, Brazil}

\author{Rafael Furlan de Oliveira\orcid{0000-0001-8980-3587}}
\affiliation[CNPEM]{Brazilian Nanotechnology National Laboratory, Zip Code 13083-970, Campinas, SP, Brazil}

\author{Rodrigo B. Capaz\orcid{0000-0001-5770-5026}}
\affiliation[CNPEM]{Brazilian Nanotechnology National Laboratory, Zip Code 13083-970, Campinas, SP, Brazil}
\alsoaffiliation[UFRJ]{Instituto de Física, Universidade Federal do Rio de Janeiro, Rio de Janeiro, RJ 21941-972, Brazil}

\author{Adalberto Fazzio\orcid{0000-0001-5384-7676}}
\email{adalberto.fazzio@lnnano.cnpem.br}
\affiliation[CNPEM]{Brazilian Nanotechnology National Laboratory, Zip Code 13083-970, Campinas, SP, Brazil}
\alsoaffiliation{Ilum School of Science, CNPEM, Zip Code 13083-970, Campinas, SP, Brazil}

\title{The role of functional thiolated molecules on the enhanced electronic transport of interconnected \ce{MoS2} nanostructures}

\abbreviations{DFT, PBE, BZ, VASP, vdW}


\date{\today}

\begin{document}

\maketitle

\begin{abstract}
Molecular linkers have emerged as an effective strategy to improve electronic transport properties on solution-processed layered materials via defect functionalization. However, a detailed discussion on the microscopic mechanisms behind the beneficial effects of functionalization is still missing. Here, by first-principles calculations based on density functional theory, we investigate the effects on the electronic properties of interconnected \ce{MoS2} model flakes systems upon functionalization with different thiol molecule linkers, namely thiophenol, 1,4-benzenedithiol, 1,2-ethanedithiol, and 1,3-propanedithiol. The bonding of benzene- and ethane-dithiol bridging adjacent armchair \ce{MoS2} nanoflakes leads to electronic states just above or at the Fermi level, thus forming a molecular channel for electronic transport between flakes. In addition, the molecular linker reduces the potential barrier for thermally activated hopping between neighboring flakes, improving the conductivity as verified in experiments. The comprehension of such mechanisms helps in future developments of solution-processed layered materials for use on 2D electronic devices.
\end{abstract}

\section{Introduction}

Two-dimensional (2D) materials have been the focus of intensive theoretical and experimental research due to their distinctive and unique physicochemical properties, which can be widely exploited in the development of new technologies. Transition metal dichalcogenides (TMD’s) have attracted attention among these materials because of their interesting  semiconducting properties, which hold tremendous potential for applications in catalysis\cite{Voiry_6197_2016,Tsai_15113_2017}, (opto)electronics\cite{Cheng_5590_2014,Cui_534_2015,Wang_699_2012}, energy storage\cite{Du_1106_2010,Chang_4720_2011}, among others\cite{Radisavljevic_9934_2011,Jariwala_1102_2014,Nguyen_6225_2015,Ferrari_4598_2015,Kelly_217_2022,Ippolito_2100122_2022}.

One of the major methods for producing large amounts of TMDs is through the so-called liquid-phase exfoliation (LPE) technique\cite{Ippolito_2021,Ippolito_2100122_2022}, where bulk crystals are exfoliated in a given solvent via mechanical energy transfer (e.g. assisted by sonication), which disrupt the van der Waals interactions within the layered material. The solution processing of TMDs allows one to produce thin-films and coatings thereof employing a variety of scalable deposition techniques, which is very appealing for printable and large area electronics\cite{Bonaccorso_6136_2016}. However, structural defects in solution processed TMDs can be a limiting factor for their electronic properties, as the exfoliation process can increase the density of structural defects, e.g. sulfur vacancies in metal dissulfides (\ce{MS2})\cite{Sim_12115_2015,Tsai_15113_2017,Backes_7050_2019}. Although structural defects in TMDs are important catalytic sites for hydrogen generation reactions, they act as charge scattering centers that may jeopardize the carrier transport characteristics TMD-based films in (opto)electronic applications\cite{Ippolito_2021,Kelly_217_2022,Ippolito_2100122_2022}.

To reduce the deleterious effects of structural defects in TMDs, chemical functionalization with molecules represents an interesting and versatile approach\cite{Bertolazzi_1606760_2017,Bertolazzi_6845_2018}. Regarding \ce{MS2} and more specifically \ce{MoS2} flakes, many studies have explored the repairing (or healing) of defects via functionalization by thiolated molecules on its basal plane\cite{Moses_104709_2009,Nguyen_6225_2015,Bertolazzi_1606760_2017,Li_10501_2017}.

Healing of sulfur vacancies in \ce{MS2} is expected to partially restore the crystalline structure of the material, thereby reducing charge scattering centers. However, for solution-processed \ce{MS2} the poor interflake connectivity still plays a major and deleterious role on the long-range charge transport within thin-films and related device applications\cite{Ippolito_2021}. Recently, Ippolito et al\cite{Ippolito_2021}. reported an innovative strategy that employs dithiolated molecules to simultaneously heal the sulfur vacancies in \ce{MoS2} and to bridge adjacent flakes. By using 1,4-benzenedithiol (BDT) molecules they functionalized solution-processed \ce{MoS2} films, which exhibited superior electrical properties, such as a 1-order of magnitude increased charge carrier mobility.

\begin{figure}
    \centering
    \includegraphics[width=0.45\textwidth]{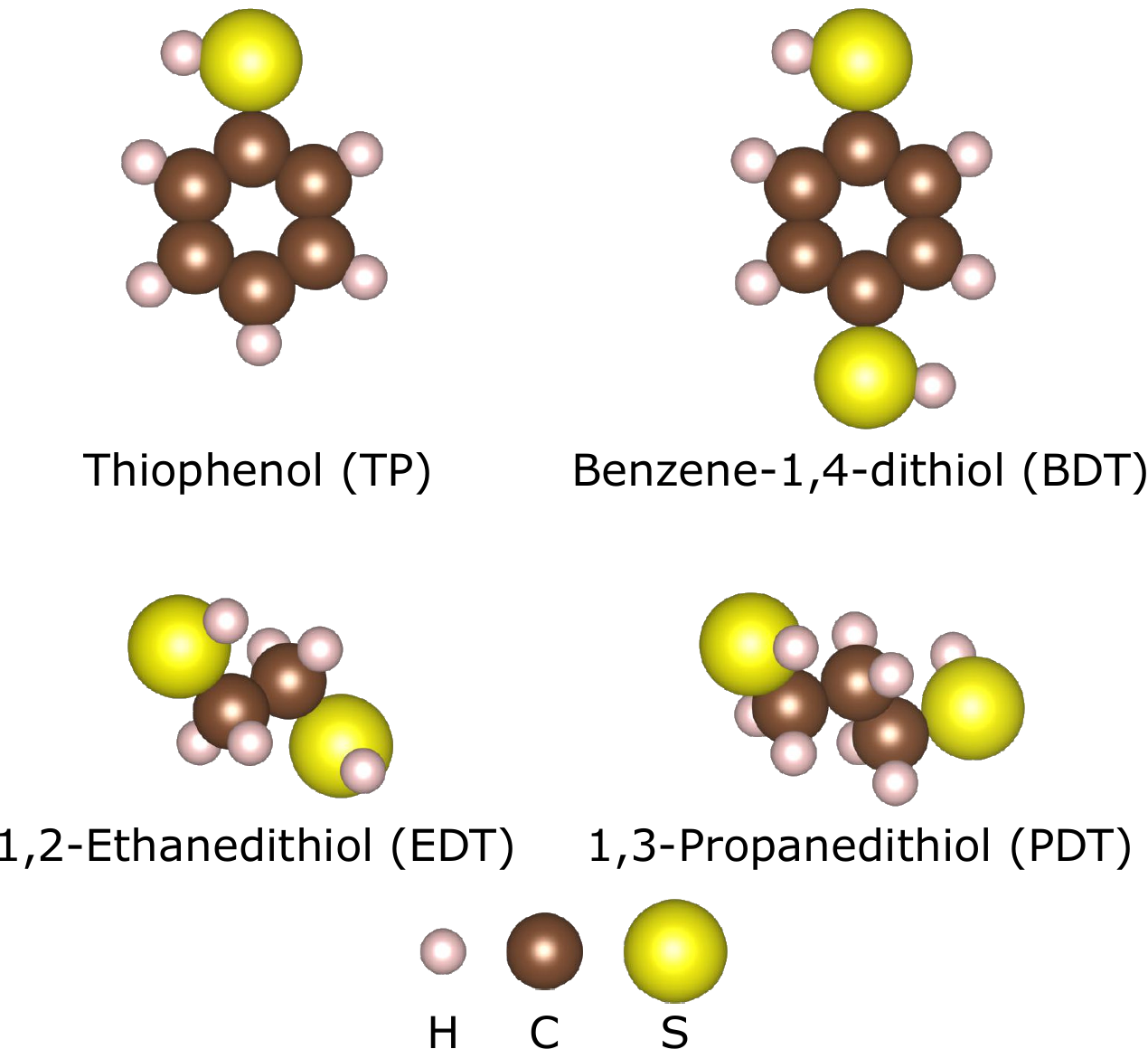}
    \caption{Molecular structure of the studied linkers between adjacent \ce{MoS2} ribbons. \ce{H}, \ce{C}, and \ce{S} are represented by white, brown, and yellow atoms, respectively, as indicated.} 
    \label{fig:linkers}
\end{figure}

In this paper, we shed light on the role of such functional dithiolated molecular linkers on the properties of interconnected \ce{MoS2} flakes by analyzing the modifications of eletronic properties due to thiol-molecules bridging of \ce{MoS2} nanoribbons. By {\it ab initio} calculations, we have studied the structural cohesion of \ce{MoS2} nanoribbons model by the thiolated molecules, namely thiophenol (TP),  1,4-benzenedithiol (BDT), 1,2-ethanedithiol (EDT), and 1,3-propanedithiol (PDT) (See Figure~\ref{fig:linkers}). As experimentally demonstrated by Ippolito {\it et al.}\cite{Ippolito_2021}, BDT molecules are endowed of two peripheral thiol groups - that are ideal to bridge adjacent flakes - and an electron-delocalized core to improve the conducitivity of bridged flakes. TP molecules, in turn, are BDT analogues possesing a single thiol group, making them healer-only molecule. EDT and PDT are dithiolated bridging molecules possesing an alkylated core of \num{2} and \num{3} single-bonded carbon atoms, respectively. The bonding of BDT and EDT on armchair \ce{MoS2} nanoribbons gives rise to dispersive electronic states just above the Fermi level, thus creating a molecular channel for electronic transport along the bridging direction for $n$-doping conditions. In contrast, monothiolated TP does not bridge adjacent flakes, as expected, and \ce{MoS2} nanoribbons bridged by PDT presents an insulating behavior. The understanding of such mechanism could be helpful to guide future experiments, particularly, for applications on 2D systems devices through the engineering of their electronic and structural properties.

\begin{figure*}
    \centering
    \includegraphics[width=0.9\textwidth]{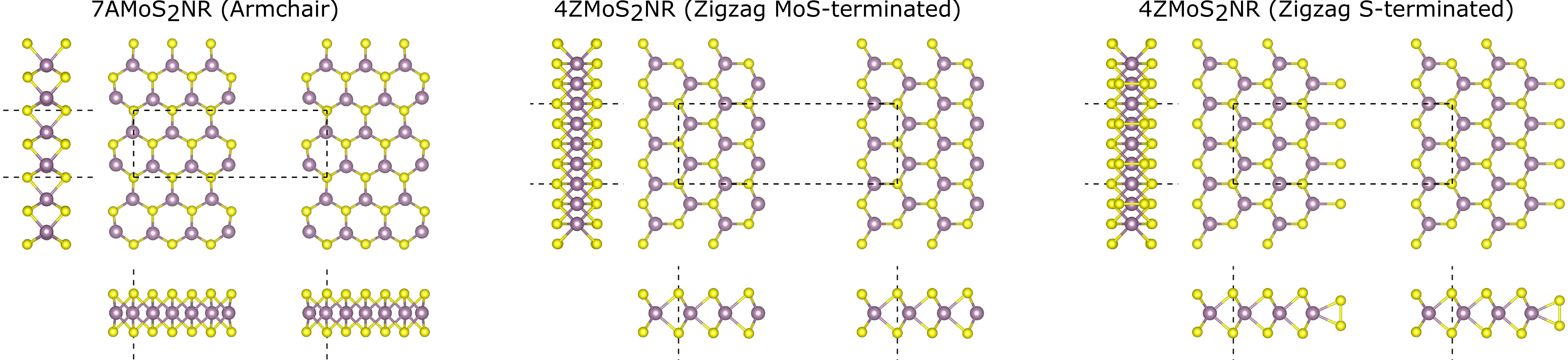}
    \caption{Relaxed \ce{MoS2} pristine nanoribbons with armchair, A\ce{MoS2}NR (left), and zigzag, Z\ce{MoS2}NR, termination (mix- and \ce{S}-terminated).}
    \label{fig:ribbons}
\end{figure*}

\section{Computational details}

We performed spin-polarized calculations based on density functional theory (DFT)\cite{Hohenberg_B864_1964,Kohn_A1133_1965} within the semi-local exchange-correlation functional proposed by Perdew--Burke--Ernzerhof (PBE)\cite{Perdew_3865_1996}. The the long range dispersion van der Waals (vdW) interactions\cite{Grimme_154104_2010} were taking into account through the well-known pairwise D\num{3} correction proposed by Grimme\cite{Grimme_154104_2010,Grimme_1456_2011}, in which an additive vdW energy correction, $E_{energy}^{\text{vdW}}$, is added to the DFT-PBE total energy, $E_{tot}^{\text{DFT}}$, i.e., $E_{tot}^{\text{DFT+vdW}} = E_{tot}^{\text{DFT}} + E_{energy}^{\text{vdW}}$\cite{Moellmann_8500_2010,Reckien_2023_2012}.

For the total energies, $E_{tot}^{\text{DFT+vdW}}$, the electron-ion core interactions were considered within the the projector augmented wave (PAW) method\cite{Blochl_17953_1994,Kresse_1758_1999}, as implemented in the  Vienna {\it Ab-Initio} Simulation Package (VASP)\cite{Kresse_13115_1993,Kresse_11169_1996}. For all calculations, the cutoff energy for the plane-wave expansion of the Kohn--Sham orbitals was set to \SI{520}{\electronvolt}, with a total energy convergence parameter of \SI{e-6}{\electronvolt}, while the equilibrium structures were reached once the atomic forces on every atom were smaller than \SI{0.001}{\electronvolt\per\angstrom}.

For total energy calculations, and additional analyses, we applied a \num{25} \textbf{k}-point mesh density for the Brillouin zone (BZ) integrations, which provided a \num{2x5x1} \textbf{k}-point mesh with a total of \num{6} \textbf{k}-points. On the other hand, to obtain accurate density of states (DOS) calculations, higher \textbf{k}-point density were required, i.e., \num{50} \textbf{k}-point mesh density. For the gas-phase molecules only the $\Gamma$-point was considered as there is no dispersion in the electronic states.

To shed light on the effects of thiol exposure on system composed of MoS$_2$ nanoribbons, the system is modeled using a periodic array of nanoribbons by keeping ``bulk-like" periodic boundary conditions along one direction (the vertical direction in Fig.~\ref{fig:ribbons}) and by creating a gap between adjacent ribbons. 
We consider three types of nanoribbons, with different edge terminations, as shown in Fig. ~\ref{fig:ribbons}: One armchair-edge nanoribbon (A\ce{MoS2}-NR) and two zigzag-edge nanoribbons (Z\ce{MoS2}-NR), such that the molecules were exposed to different chemical environments. In the case of Z\ce{MoS2}-NR, yet two different terminations were considered: A mixed-termination ribbon, where a \ce{Mo}-terminated edge faces a \ce{S}-terminated edge across the gap (mix-term), and a \ce{S}-terminated ribbon (\ce{S}-term). All structures are shown in Figure~\ref{fig:ribbons}.

The unit cell parameters were fixed and the atomic positions were allowed to relax.
The molecules were placed at the periodic gaps between the  nanoribbons, thus providing interconnection to the whole system. In addition, the \ce{H}'s bonded to the \ce{S} at the molecule endings were removed and the molecule \ce{S} atoms replaced \ce{S} atoms from the \ce{MoS2}-NR edges (one or two \ce{S} atoms, either on both edges or on a single edge, depending on the case considered), as can be seen in Figure~\ref{fig:bdstr}. 

\section{Results}

\subsection{Energetic and geometric properties}

\begin{table}[ht!]
    \caption{Formation energy $E_f$ (eV) and binding energy $E_b$ (eV~\AA$^{-1}$) as calculated through Equation~\ref{eq:form-en} and~\ref{eq:bind-en} for both armchair- and zigzag-edge termination. $E_{f}^i$ and $E_{b}^i$ with $i=mix,\ce{S}$ refer to mix- and \ce{S}-terminated zigzag edges, respectively.}
    \label{table:stability}
    \resizebox{\columnwidth}{!}{%
    \begin{tabular}{lccccccc}\toprule
              & \multicolumn{2}{c}{A\ce{MoS2}-NR} && \multicolumn{4}{c}{Z\ce{MoS2}-NR}\\ \cmidrule{2-3}\cmidrule{5-8}
        Linker & $E_{f}$ & $E_{b}$ && $E_{f}^{mix}$ & $E_{b}^{mix}$ & $E_{f}^{\ce{S}}$ & $E_{b}^{\ce{S}}$\\
\midrule
       BDT    &\tablenum{-0.56}   &\tablenum{-0.45} &&\tablenum{-3.98} &\tablenum{-0.93} &\tablenum{-1.77} &\tablenum{-0.58}\\
       EDT    &\tablenum{-0.18}   &\tablenum{-0.62} &&\tablenum{-3.97} &\tablenum{-1.13} &\tablenum{-0.74} &\tablenum{-0.62}\\
       PDT    &\tablenum{-0.61}   &\tablenum{-0.70} &&\tablenum{-3.96} &\tablenum{-1.13} &\tablenum{-1.27} &\tablenum{-0.71}\\
       TP     &\tablenum{-0.90}   &\tablenum{-0.40} &&\tablenum{-3.08} &\tablenum{-0.69} &\tablenum{-1.48} &\tablenum{-0.43}\\ \bottomrule
    \end{tabular}
    }
\end{table}

For all thiol linked nanoribbon configurations, we calculate the formation energies as
\begin{equation}
    E_{f} = (E_{total}^{linker+ribbon} + xE_{g}^{\ce{H2 ^}}) - (E_{total}^{ribbon+V_{\ce{S}}^{0}} + E_{g}^{linker})
    \label{eq:form-en}
\end{equation}
in which $E_{total}^{linker+ribbon}$ and $E_{g}^{\ce{H2 ^}}$ are the equilibrium total energy of the \ce{MoS2}-NR linked by the molecule and the \ce{H2} gas phase energy, respectively. $E_{total}^{ribbon+V_{\ce{S}}^{0}}$ and $E_{g}^{linker}$ is the defected nanoribbon system energy with \num{1} or \num{2} sulfur vacancies ($V_{\ce{S}}$), and the molecule gas-phase energy, respectively. Therefore, it is assumed that the linkers will passivate S vacancies at the edges. The first term in Equation~\ref{eq:form-en} is the energy of the final state of the whole system, where $x$ hydrogen molecules per unit cell are released after the molecules link the nanoribbons. Note that $x$ need not to be integer: For instance, in the particular case of a single TP linker, $x=1/2$ since only one \ce{H} atom will be released. The second term is the energy of the initial state, in which the nanoribbon with some vacancy defects is exposed to a thiolated molecule that will react with such defects and other available ribbon surface and edge active sites. Thus, negative (positive) values indicate whether the reaction is favorable (unfavorable). For all explored systems the molecule incorporation is an exothermic process (see Table~\ref{table:stability}), indicating the formation of the NR/thiol/NR bridge.

Additionally, we evaluate the molecules binding energies as given by
\begin{equation}
    E_{b} = E_{total}^{linker+ribbon} - (E_{total}^{ribbon+V_{\ce{S}}^{0}} + E_{g}^{linker-H}),
    \label{eq:bind-en}
\end{equation}
in which the last term represents the gas-phase energy of the molecule without the \ce{H} bonded to the \ce{S} atoms. Here the bridge cohesion is evaluated (see Table~\ref{table:stability}), with the PDT and EDT offering the strong inter-flake connection reaching values up to $-1.13$\,eV/{\AA}.

\subsection{Electronic Structure}

\begin{figure*}
    \centering
    \includegraphics[width=0.8\textwidth]{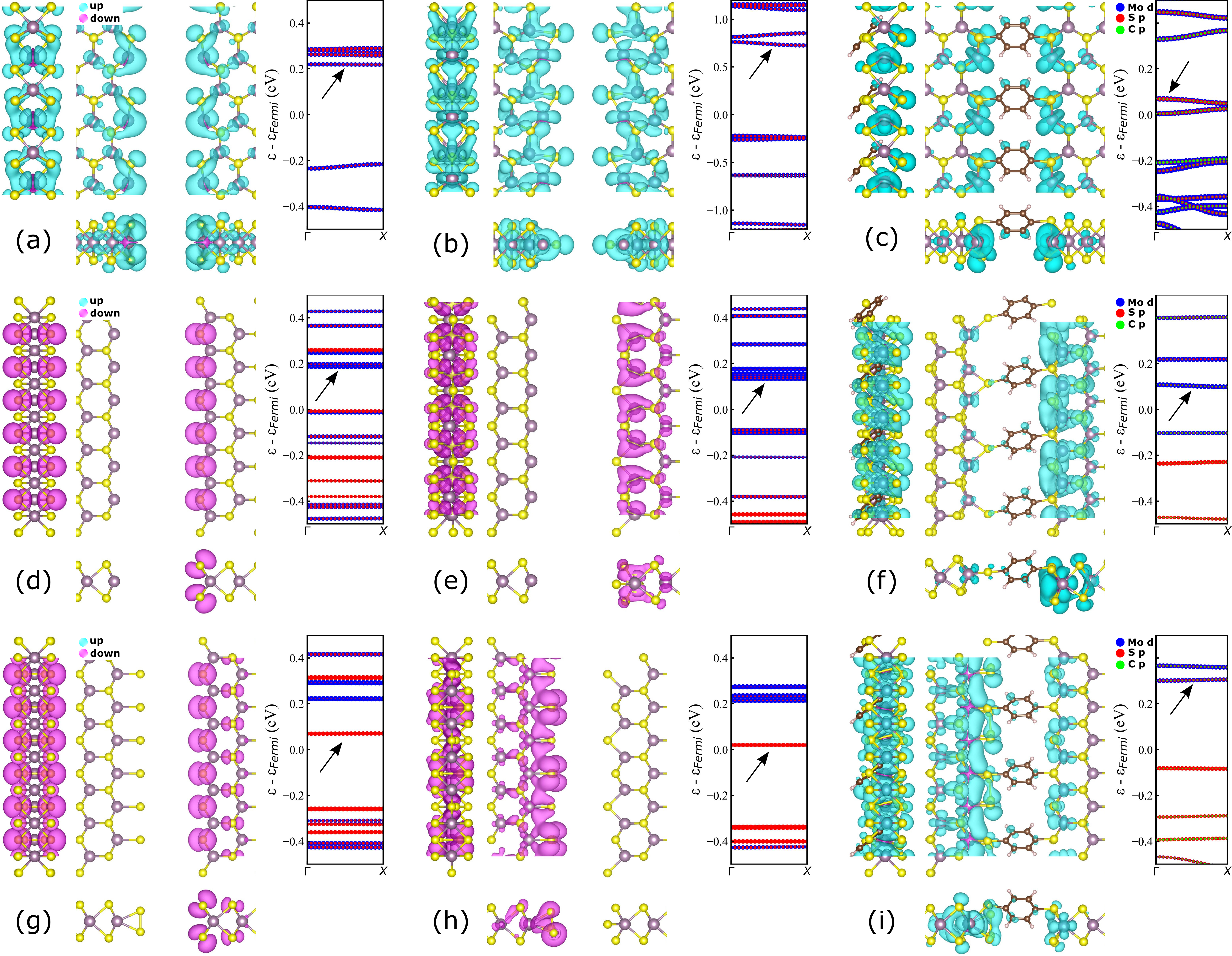}
    \caption{Armchair- and zigzag-edge nanoribbons band structure comparison. For each system, namely pristine ((a)-(d)-(g)), vacancy ((b)-(e)-(h)), and BDT bonded ((c)-(f)-(i)), we present the band structure and the projected local density of states to the band indicated by the black arrow.} 
    \label{fig:bdstr-ldos}
\end{figure*}

In this section, we discuss the effects of thiolated molecules on the band structures of \ce{MoS2}-NR.
{We focus on providing a plausible theoretical explanation to the most important results of Ippolito {\it et al.}~\cite{Ippolito_2021}. To this end, we recall a few of those results:}

{(i) Samples are composed of very large flakes (between 250 nm and 400 nm lateral size and between 13 to 23 layers). Therefore, it is expected that electronic transport is dominated by bulk-like electronic states rather than edge states.}

{(ii) Electronic transport is $n$-type and thermally activated, with Arrhenius activation energies decreasing from 0.5 eV to 0.3 eV upon insertion of BDT molecules. {Furthermore, barriers decrease with the applied bias.}}

{(iii) BDT molecules contribute to improve transport properties not only by healing defects (i.e. vacancies) in the basal plane but also - and primarily - by providing molecular links between MoS$_2$ layers.}

{The main conjecture is that such molecular linkers provide an efficient interlayer coupling between 2D bulk-like electronic states in the conduction band ($n$-doping). Therefore, we initially search for the best model system to test this conjecture.} From the literature,  \ce{MoS2}-NR are known to present different behaviors depending on their edge terminations. For instance, zigzag-edge presents metallic character, whereas the armchair one is semiconducting\cite{Botello-Mendez_325703_2009,Ataca_3934_2011,Pezo_11359_2019}. 

{The result of our initial exploration is displayed in Figure~\ref{fig:bdstr-ldos}}. Since we are primarily interested in the electronic transport across nanoribbons (interflake transport), we focus our band structure plots on the $\Gamma-X$ direction, where $x$ is perpendicular to the edges (horizontal direction in all figures). {We also focus on the BDT molecule as a linker. Results for other molecules and exploring additional directions in the Brillouin Zone are presented in the Supporting Information.} Our results confirm the metallic behavior of the pristine zigzag-edge ribbons with a magnetic polarization, while the armchair presents a band gap with a non-magnetic ground state. We also investigate the effects of sulfur vacancies in the band structure: For armchair systems, we observe an increase in the band gap and a small dispersion of the band close to the Fermi level along with the $\Gamma-X$ direction; whereas for zigzag ribbons the presence of sulfur vacancies did not change the metallic behavior from the pristine systems, and no dispersion could be observed along the $\Gamma-X$ direction as one can see in Figure~\ref{fig:bdstr-ldos}.
{Naively, one would think that the metallic behavior of the zigzag configuration would be better to explore for electronic transport purposes. However, even if the bands are close to the Fermi level, they correspond to very localized edge states and they should not contribute significantly to electron transport in very large flakes (as seen from their flatness along the $\Gamma-X$ direction). On the other hand, our armchair NR system seems to be very convenient to model the coupling between bulk states, as it does not introduces edge states in the gap. Therefore the effects of molecular linkers on the $n$-doped conduction bands can be directly probed.} 

{Notice that the different nanoribbon structures behave quite differently in terms of the spatial distribution of the electronic density associated to conduction band states (projected local density of states). As shown in Figure~\ref{fig:bdstr-ldos}, while pristine and defected armchair-edge NRs present an electronic density typical of bulk states (distributed over the entire NR width), in the zigzag ones the electronic density is quite localized at one of nanoribbon edges, independent of the zigzag edge termination. After molecular functionalization, the qualitative picture remains the same: For armchair edges, conduction band states are delocalized over the entire system, including contributions from the BDT molecule, whereas for zigzag edges the charge density is still more localized at one side than the other.} 
Additional analyses can be found in the Supporting Information (SI).

\begin{figure*}
    \centering
    \includegraphics[width=0.8\textwidth]{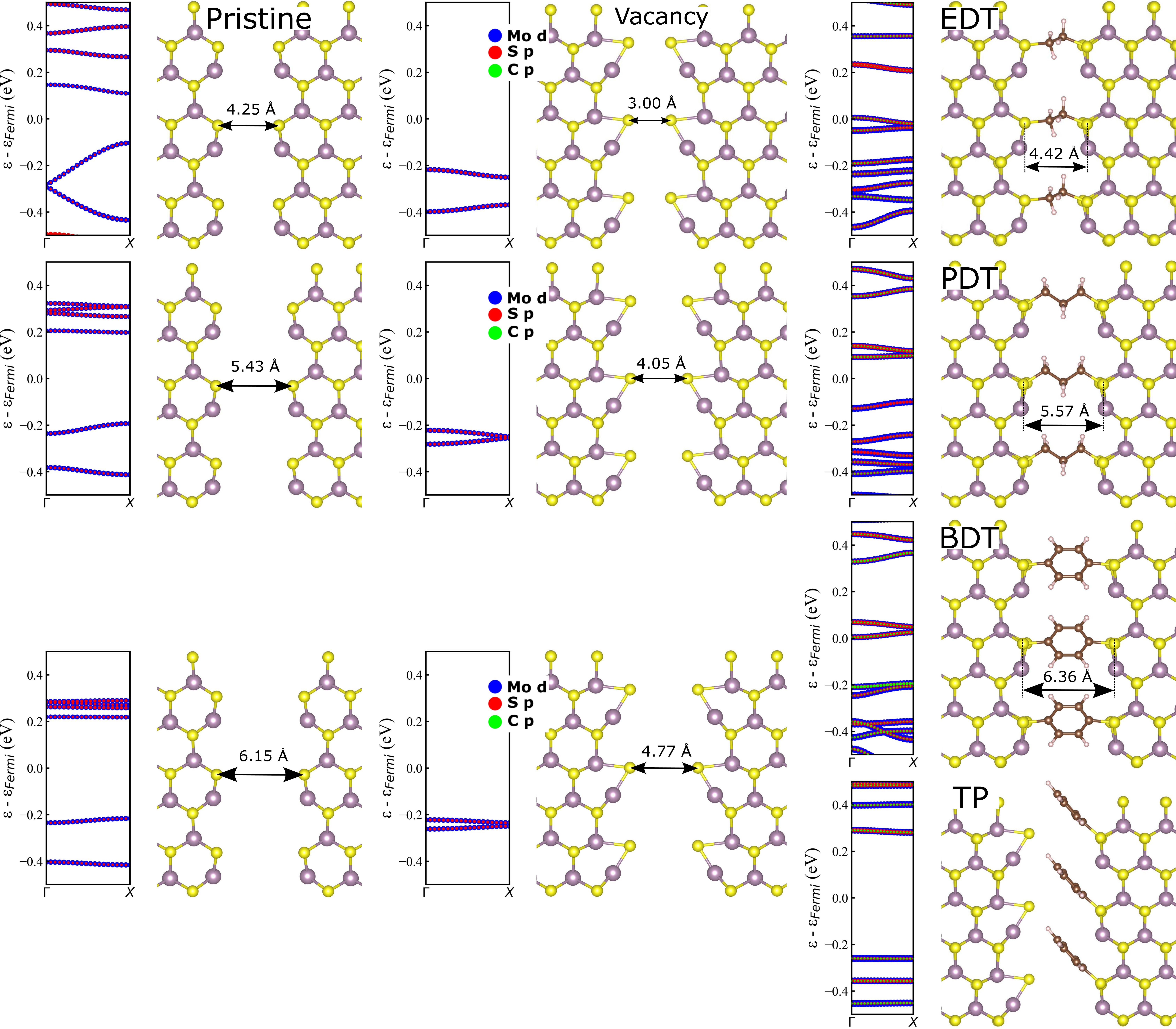}
    \caption{Band structures of \ce{MoS2} ribbons. Left panels are the pristine systems, Middle panels are the vacancy systems, and Right panels are the molecularly bonded systems.} 
    \label{fig:bdstr}
\end{figure*}

{As expected, the observed differences in the localization properties of conduction band states have direct implications in the bandwidth of these states. As seen in Figure~\ref{fig:bdstr-ldos}, and also as discussed in further detail in Figure~\ref{fig:bdstr}, in the case of  BDT and EDT, for armchair edges, the molecular linker will contribute to more disperse bands. These results not only suggest that armchair-edge NRs are the most suitable model structures to investigate the coupling between bulk states in neighboring flakes induced by molecular linkers, but also strongly indicate that molecules such as BDT have the ability to provide a resonant coupling between these states, which result in a molecular channel that facilitates interflake conduction. For these reasons, in the remainder of this work} we focus our attention on the armchair-edge nanoribbon (A\ce{MoS2}-NR).  Additional results for other structures are presented in the Supporting Information. 

Besides ribbon edge termination, at least two features are observed to influence the band structure of the systems: 

(i) molecular linker: As reported by Ippolito {\it et al.}, \cite{Ippolito_2021}, after BDT exposure there is an enhancement of the electrical conductivity of \ce{MoS2} networks, which is attributed to a molecule-mediated link between adjacent flakes. Our band structure calculations reveal a more disperse behavior of the bands along the $\Gamma-X$ path as the \ce{MoS2}-NR is exposed to thiolated molecules, if compared to both pristine and defected structures. 
Among all molecules considered, the most promising results occur for EDT and BDT molecules (the latter in agreement with experiment\cite{Ippolito_2021}). {BDT linkers give rise to dispersive bands just above the Fermi level, indicating that $n$-type conductance can be easily promoted by appropriated doping. In the case of EDT, dispersive bands appear right at the Fermi level, indicating metallic behavior. On the other hand, large band gaps remain for TP and PDT linkers indicating a more insulating behavior.} Therefore, we have two linear alkyl molecules (EDT and PDT) which can both provide a network formation between adjacent nanoribbons, but whose effects on the electronic structure are slightly different. The better performance of EDT over PDT may be  related to its small chain length, which allows for a small inter-flakes distance (about \SI{4.4}{\angstrom} for EDT, and \SI{5.6}{\angstrom} for PDT after functionalization) and an increased interaction between the nanoribbons edges. On the other hand, both TP and BDT also provide larger inter-flake distances, but their $\pi$-conjugated characteristic with electron delocalization can contribute in the electronic transport between adjacent flakes after functionalization, {as long as these states are resonant with ``bulk-like" conduction band states (for the case of $n$-doping.) This seems to be the case only for BDT, as TP does not allow network formation because it makes covalent bonding only to one of the two edges, and therefore it does not contribute to bridge possible pathways for electron propagation.}

\begin{figure*}[h!]
    \centering
    \includegraphics[width=0.80\textwidth]{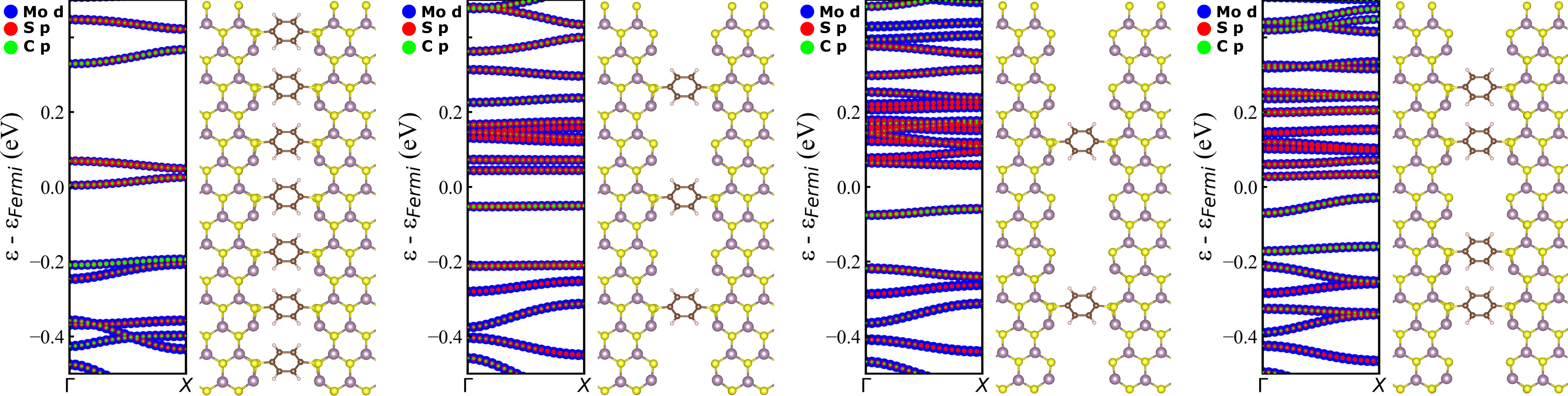}
    \caption{Effect of BDT molecule concentration. From left to right, the linear density of BDT molecules along the edges are \SIlist[list-final-separator={, and }]{0.18;0.09;0.06;0.12}{\per\angstrom}, respectively.}
    \label{fig:moldist}
\end{figure*}

(ii) the molecule concentration and distribution along nanoribbon edge: We analyzed the impact of molecular concentration and distribution along the nanoribbon edge in the electronic structure of \ce{MoS2}-NR by increasing the $2D$ unit cell length along the $y$-direction, such that molecules are farther apart from each other. In this analysis, we only considered the BDT molecule. 
As one can see in Figure~\ref{fig:moldist}, as the number of BDT molecules over the boundary decreases towards low concentrations, the band gap undergoes a slight increase and the bands dispersion are less pronounced than in the high concentration limit. Besides, we also studied the effect of adding molecules in pairs (grouped), but far apart (right corner of Figure~\ref{fig:moldist}). When the molecules are grouped, despite of the distance from the next pair of molecules, we can see bands close to the Fermi level, as in the high-concentration case. Furthermore, the gap decreases and the bands close to the $E_F$ become more disperse again. 

\FloatBarrier
\section{Thermally Activated Transport and Electrostatic Potential Energy Barrier}

\begin{figure}[ht!]
    \centering
    \includegraphics[width=0.4\textwidth]{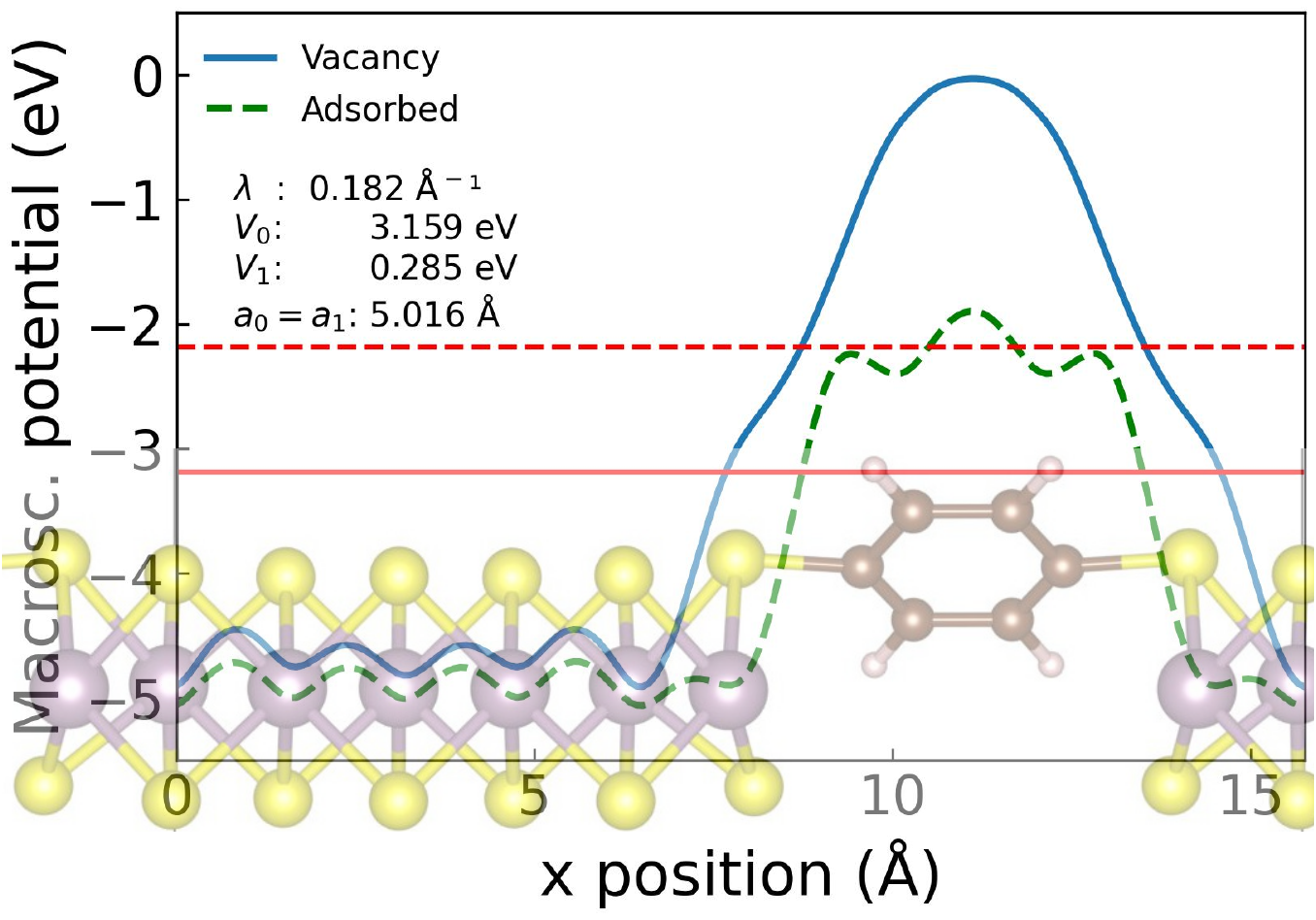}
    \caption{Electrostatic potential energy barrier for \ce{MoS2}-NR linked by BDT molecule. The red lines represent the Fermi level of vacancy (solid line) and bonded (dashed line) systems which were used as reference to calculate the energy barriers $V_0$ and $V_1$, respectively.}
    \label{fig:barrier}
\end{figure}

{As discussed previously, experimental results \cite{Ippolito_2021} suggest that electronic conduction in both pristine and molecular-linked MoS$_2$ films is $n$-type and thermally activated. On the other hand, our calculations indicate that BDT-linked flakes develop electronic states just above the Fermi level with some energy dispersion, indicating that BDT molecules provide a resonant molecular pathway for electrons between two neighboring flakes. Based on this information, a number of possible conduction mechanisms\cite{Schmidt_7715_2015,Kelly_217_2022} can be conjectured and analyzed:}

{(i) {\it Metallic conduction}: If the Fermi level is positioned (by doping) at the resonant BDT-MoS$_2$ band, metallic conduction could in principle occur through this band. However, the temperature dependence of metallic conduction does not correspond to thermal activation. Therefore, this possibility can be ruled out.}

{(ii) {\it Activation from defect states}: As usual in semiconductors, $n$-type conduction can be activated by the thermal release of carriers from localized defect states to the conduction band. However, our calculations indicate that the BDT-MoS$_2$ states just above the Fermi level are not defect-localized, but rather extended through the whole system.}

{(iii) {\it Thermally-activated hopping}: Even when there is a non-zero density-of-states at the Fermi level, disorder and Coulomb effects can lead to localization and hopping conduction. Thermally-activated hopping over a potential barrier (Schottky mechanism) can take place under these conditions. In this case, the effect of molecular linkers would consist of decreasing the energy barrier for the electronic hopping between neighboring flakes. 
}

We test this conjecture by investigating the electrostatic potential changes of \ce{MoS2}-NR in the presence of thiolated molecules. 
In Figure~\ref{fig:barrier}, we show the macroscopic average of the electronic potential energy in the case of BDT linker, before and after bonding. The same analysis was done for all molecules and nanoribbons edges and is presented in the SI. {The barrier for Schottky emission is given by the energy difference between the Fermi level (horizontal lines) and the peak in the potential energy \cite{Funck_045116_2019}. Figure~\ref{fig:barrier} describes the situation of maximum BDT concentration between adjacent edges and shows a substantial decrease of the tunneling barrier upon molecule linking. As seen in the figure, the barrier decreases from 3.2 eV (for tunneling through vacuum) to 0.3 eV (for tunneling through the BDT molecule}. Interestingly, this value agrees very well with the experimental findings, in which energy barriers about \SI{0.36}{\electronvolt} were observed by using a saturated BDT solution within a soaking process\cite{Ippolito_2021}. Thus, we have shown that the molecular linkers play a fundamental role in decreasing the tunneling barrier for electron transport between adjacent \ce{MoS2} flakes.

\FloatBarrier
\section{Summary}

Through a density functional theory investigation, we have studied the effects of thiolated molecular linkers on the electronic properties of MoS$_2$ flakes (here modelled by periodic sequence of adjacent \ce{MoS2} nanoribbons). We found that BDT and EDT molecule mediated-bridging of \ce{MoS2}-NR give rise to bands just above the Fermi level of \ce{MoS2}, thus appropriate to $n$-type conduction. The associated electronic states are extended throughout the ribbon and resonant with molecular states, thus providing a electronic conduction pathway between neighboring flakes. Coulomb effects or disorder, always present in real samples, may lead to hopping conduction through localized states derived from these bands. Calculated energy barriers for thermally-activated hopping decrease upon molecule insertion and agree well with experimental measurements \cite{Ippolito_2021}, suggesting a plausible mechanism for the improvement of \ce{MoS2}-NR electronic properties, with direct impact on its electronic conductivity properties. 
The understanding of such mechanism could help to guide future experiments, and development and designing of $\num{2}D$ electronic devices, by improving and engineering their electronic properties.

\begin{suppinfo}
Extra data and analysis are provided in the Supporting Information.
\end{suppinfo}

\begin{acknowledgement}
We thank the S\~ao Paulo Research Foundation, FAPESP (Grant 2020/14067-3).
\end{acknowledgement}

\FloatBarrier
\bibliography{jshort_25set2017,boxref_18oct2021}

\end{document}